\documentclass[12pt]{article}
\usepackage{a41}
\usepackage{epsfig}
\usepackage{cite}
\usepackage{amsmath}
\usepackage{amssymb}
\usepackage{float}
\usepackage{verbatim}
\usepackage{amsxtra}
\usepackage{afterpage}

\newcommand{\N}{\nonumber} 
\newcommand{\ep}{\varepsilon}

\newcommand{\bea}{\begin{eqnarray}}
\newcommand{\bq}{\begin{equation}}
\newcommand{\eea}{\end{eqnarray}}
\newcommand{\eq}{\end{equation}}

\newcommand{\gsim}{\raisebox{-0.07cm   }
{$\, \stackrel{>}{{\scriptstyle\sim}}\, $}}

\newcommand\MSbar{$\overline{\mbox{MS}}$}
 \newcommand{\MS}{\overline{{\sf MS}}}

\newcommand\be{\begin{eqnarray}}
\newcommand\ee{\end{eqnarray}}

\newcommand\Li{{\rm Li}}
\newcommand\HA{{\rm H}}

\newcommand{\Ahathat}{\hat{\hat{A}}}
\begin{document}
\noindent
\sloppy
\thispagestyle{empty}
\begin{flushleft}
DESY 12-055 \hfill %{\tt arXiv:12xx.xxx [hep-ph]}
\\
DO-TH-12/13
\\
SFB/CPP-12-28\\
LPN 12-052\\
May 2012
\end{flushleft}
%
%\setcounter{page}{0}
% 1
%\mbox{}
\vspace*{\fill}
\hspace{-3mm}
{\begin{center}
{\LARGE \bf
The {\boldmath $O(\alpha_s^3 n_f T_F^2 C_{A,F})$} Contributions to the}

\vspace*{2mm}
{\LARGE \bf Gluonic Massive Operator Matrix Elements}

\end{center}
}

\begin{center}
\vspace{2cm}
\large
Johannes Bl\"umlein$^a$, Alexander Hasselhuhn$^a$, Sebastian Klein$^b$,
\newline
and Carsten Schneider$^c$
\vspace{5mm}
\\

\vspace{5mm}
\normalsize
{\itshape $^a$~Deutsches Elektronen--Synchrotron, DESY,\\
Platanenallee 6, D--15738 Zeuthen, Germany}
\\

\vspace{5mm}
\normalsize {\itshape $^b$~Institute for Theoretical Physics E, RWTH Aachen University, 
D--52056 Aachen, Germany} \\ 

\vspace{5mm}
\normalsize {\itshape $^c$~Research
Institute for Symbolic Computation (RISC),\\ Johannes Kepler
University, Altenbergerstra\ss{}e 69, A-4040 Linz, Austria}
\\

%\today
\end{center}

\vspace*{\fill} %
%%%%%%%%%%%%%%%%%%%%%%%%%%%%%%%%%%%%%%%%%%%%%%%%%%%%%%%%%%%%%%%%%%%%%%%%
\begin{abstract}
\noindent
The $O(\alpha_s^3 n_f T_F^2 C_{A,F})$ terms to the massive gluonic operator 
matrix elements are calculated for general values of the Mellin variable $N$. 
These twist-2 matrix elements occur as transition functions in the variable 
flavor number scheme at NNLO. The calculation uses sum-representations in 
generalized hypergeometric series turning into harmonic sums. The analytic 
continuation to complex values of $N$ is provided.
\end{abstract}
%%%%%%%%%%%%%%%%%%%%%%%%%%%%%%%%%%%%%%%%%%%%%%%%%%%%%%%%%%%%%%%%%%%%%%%%

\vspace*{\fill} 

\newpage

\vspace*{1cm}\noindent
%%%%%%%%%%%%%%%%%%%%%%%%%%%%%%%%%%%%%%%%%%%%%%%%%%%%%%%%%%%%%%%%%%%%%%%
\section{Introduction}
\label{sec:1}
%%%%%%%%%%%%%%%%%%%%%%%%%%%%%%%%%%%%%%%%%%%%%%%%%%%%%%%%%%%%%%%%%%%%%%%

\vspace*{1mm}
\noindent
Heavy quark contributions to the deep inelastic scattering structure functions 
play a crucial role in the QCD analyses to determine the parton distribution
functions and the strong coupling constant $\alpha_s(M_Z^2)$ in a consistent
manner, cf.~\cite{Alekhin:2012ig}. The heavy flavor corrections were calculated
at NLO in semianalytic form in~\cite{HEAV2}\footnote{A fast and precise numerical
implementation in Mellin space has been given in~\cite{Alekhin:2003ev}.}. To avoid 
contributions of higher twist the analysis has to be restricted to large enough 
values of $Q^2$. It has been shown in~\cite{BMSN96} that for $Q^2 \gsim 10~m^2$, with 
$m$ the heavy quark mass, the heavy flavor contributions to the structure function 
$F_2(x,Q^2)$ are rather accurately described using the asymptotic representation in 
which all power corrections $\propto (m^2/Q^2)^k, k \in \mathbb{N}_+$ are neglected. 
In this case the heavy flavor Wilson coefficients can be calculated analytically. They 
are given by convolutions of massive operator matrix elements (OMEs) and the massless 
Wilson coefficients, cf.~Ref.~\cite{BMSN96,BBK09NPB}. The massless Wilson coefficients 
are known to 3-loop order~\cite{MVV2005}. At NLO the massive OMEs were calculated in  
\cite{BMSN96,BBK07NPB,BMSN98,BBK09PLB,Buza:1996xr,Bierenbaum:2007pn,BBKS08NPB} in the 
unpolarized and polarized case, including the $O(\alpha_s^2 \ep)$ contributions, and in 
\cite{Blumlein:2009rg} for transversity. The heavy flavor corrections for charged current 
reactions are available at one loop and in the asymptotic case at two-loops 
\cite{CC,Blumlein:2011zu}. 

At 3-loop order a series of moments has been calculated for all massive 
OMEs for $N = 2 ... 10 (14)$ contributing in the fixed and variable flavor scheme, 
\cite{BBK09NPB}. The  3-loop heavy flavor corrections to $F_L(x,Q^2)$ in the 
asymptotic case were calculated in \cite{Blumlein:2006mh}. First results for general 
values of $N$ have been obtained for 
the OMEs with operator insertions on the quark lines in case for the color factors
$n_f T_F^2 C_{A,F}$ \cite{ABKSW11NPB} and 3-loop ladder topologies \cite{ABHKSW12}.
First $T_F^2 C_{A,F}$-contributions at general $N$ were calculated in 
\cite{Ablinger:2011pb} for two heavy quark lines carrying the same mass. Furthermore, 
the moments $N = 2,4,6$ in case of the OMEs contributing to the structure 
function $F_2(x,Q^2)$ with two different heavy quark masses were computed in
\cite{Ablinger:2011pb,Ablinger:2012qj}. In all the above cases the massive OMEs are 
calculated for external massless partons which are on-shell. The case of massive 
on-shell external lines has been treated in \cite{Blumlein:2011mi} recently.

In the present paper the 3-loop corrections of $O(n_f T_F^2 C_{A,F})$ to the massive 
OMEs with local operator insertions on the gluonic lines, $A_{gq,Q}$ and $A_{gg,Q}$, 
at general values of $N$ are calculated. Together with the corresponding terms with 
the insertions on the quark lines, \cite{ABKSW11NPB}, these contributions complete
all terms corresponding to the case of one massless and one massive fermion line
at 3-loop order. These matrix elements contribute to the transition functions needed
to describe the parton densities in the variable flavor number scheme (VFNS). In this
scheme it is possible to define heavy quark distribution functions assuming that there
exists only one heavy quark and all other quarks can be dealt with as massless
in the sense of an effective field theory approach. These distributions can be used
for effective calculations in some processes at hadron colliders. The picture holds to 
2-loop orders. Starting with the 3-loop corrections, 
\cite{Ablinger:2011pb,Ablinger:2012qj}, 
dia- grams containing quarks of two different masses contribute even to the universal 
corrections. Since $m_c^2/m_b^2 \approx 1/10$ is not a small number, the original 
VFNS-picture does not necessarily hold in practice. Here we deal with the  $O(n_f T_F^2 
C_{A,F})$ contributions which are in accordance with the VFNS. In Section~2 the main 
formalism is lined out. The calculation is performed in $D = 4 + \ep$ dimensions and uses 
representations in terms of generalized hypergeometric functions. They lead to multiple
sum representations, which are solved using modern summation technologies encoded in
the package {\tt Sigma} \cite{SIGMA}. The results of the calculation are given in Section~3 both 
in Mellin-$N$ and in $x$-space, and Section~4 contains the conclusions. 

\vspace*{-3mm}
%%%%%%%%%%%%%%%%%%%%%%%%%%%%%%%%%%%%%%%%%%%%%%%%%%%%%%%%%%%%%%%%%%%%%%%
\section{Parton distribution functions in the VFNS}
\label{sec:2}
%%%%%%%%%%%%%%%%%%%%%%%%%%%%%%%%%%%%%%%%%%%%%%%%%%%%%%%%%%%%%%%%%%%%%%%

\vspace{1mm}
\noindent
The neutral current Born cross section of unpolarized deep inelastic scattering (DIS) 
is given by \cite{Arbuzov:1995id}
%--------------------------------------------------------------------
\begin{eqnarray}
\label{eq:BORN}
\frac{d^2 \sigma_B^{\rm NC}}{dx dy} 
 &=& 
  \frac{2\pi \alpha^2}{xyQ^2}
  \Biggl\{
   \left[2(1-y) - 2 xy \frac{M^2}{S} + \left(1 + 4x^2 \frac{M^2}{Q^2}\right)\frac{y^2}
  {1+R(x,Q^2)}\right]{\cal F}_2(x,Q^2) 
\N\\
&&
+ xy(1-y) {\cal F}_3(x,Q^2)
  \Biggr\}~,
\end{eqnarray}
%--------------------------------------------------------------------
neglecting lepton mass contributions. Here $x$ and $y$ denote the Bjorken variables
and $-q^2 = Q^2 = xyS$, with $q^2$ the 4-momentum transfer. The structure functions
${\cal F}_i(x,Q^2)$ contain electro-weak effects due to $Z$-boson exchange and differ
for lepton and anti-lepton-nucleon scattering, cf. 
\cite{Arbuzov:1995id}, and
%--------------------------------------------------------------------
\begin{eqnarray}
R(x,Q^2) = \left(1+ 4x^2 \frac{M^2}{Q^2}\right)
\frac{{\cal F}_2(x,Q^2)}{2x{\cal F}_1(x,Q^2)} - 1~.
\end{eqnarray}
%--------------------------------------------------------------------
In the limit $M_Z^2 \gg Q^2$ the electromagnetic terms in (\ref{eq:BORN})
dominate and only the two  structure functions $F_{1,2}(x,Q^2)$ contribute, with
%--------------------------------------------------------------------
\begin{eqnarray}
2x F_1(x,Q^2) = F_2(x,Q^2) - F_L(x,Q^2)~,
\end{eqnarray}
%--------------------------------------------------------------------
where $F_L$ is the longitudinal structure function. Both structure functions contain
light and heavy quark contributions. The $y$-dependence of the differential scattering 
cross section is used to separate the structure functions \cite{FL} and allows precise
measurements of the structure function $F_2(x,Q^2)$. In the twist-2 approximation, 
referring to the fixed flavor number scheme, they are given by
%---------------------------------------------------------------------------------
\begin{eqnarray}
F_2(x,Q^2,n_f) = F_2^{m=0}(x,Q^2,n_f) + F_{2,Q}^{\rm massive}(x,Q^2,n_f,m)~.
\end{eqnarray}
%---------------------------------------------------------------------------------
Here $F_2^{m=0}(x,Q^2)$ denotes the well-known massless contribution and the massive
contribution in the presence of a single massive quark reads \cite{BBK09NPB} 
%---------------------------------------------------------------------------------
{\small
\begin{eqnarray}
\label{eqF2}
F_{2,Q}^{\rm massive}(x,Q^2,n_f,m)   
\vspace*{-2mm}
&=&
\vspace*{-2mm}   
\sum_{k=1}^{n_f} e_k^2 \Biggl\{
L_{2,q}^{\sf NS}\left(n_f,\frac{Q^2}{m^2},\frac{m^2}{\mu^2} \right)
\otimes \Bigl[f_k(x,\mu^2,n_f) + f_{\overline{k}}(x,\mu^2,n_f)\Bigr]
\N\\
&&
+{\tilde{L}}_{2,q}^{\sf PS}\left(n_f,\frac{Q^2}{m^2},\frac{m^2}{\mu^2} \right) \otimes 
\Sigma(x,\mu^2,n_f)
%\nonumber\\
%&&
+
{\tilde{L}}_{2,g}^{\sf S}\left(n_f,\frac{Q^2}{m^2},\frac{m^2}{\mu^2}
\right) \otimes G(x,\mu^2,n_f) \Biggr\}
%\end{eqnarray}\begin{eqnarray}
\nonumber\\  
&&
+ e_Q^2 \Biggl[H_{2,q}^{\sf PS}
\left(n_f,\frac{Q^2}{m^2},\frac{m^2}{\mu^2}\right) \otimes \Sigma(x,\mu^2,n_f)
\nonumber\\ &&
+ H_{2,g}^{\sf S}
\left(n_f,\frac{Q^2}{m^2},\frac{m^2}{\mu^2}\right) \otimes G(x,\mu^2,n_f)
\Biggr]~.   
\end{eqnarray}
}
%---------------------------------------------------------------------------------

\noindent
Here $f_k(x,\mu^2,n_f), \Sigma(x,\mu^2,n_f), G(x,\mu^2,n_f)$ denote the $k$th quark,
singlet-quark, and gluon densities, respectively with
%--------------------------------------------------------------------
\begin{eqnarray}
 \Sigma(x,n_f,\mu^2) 
  = \sum_{k=1}^{n_f} [f_k(x,n_f,\mu^2) + f_{\bar{k}}(x,n_f,\mu^2)]~.
\end{eqnarray}
%--------------------------------------------------------------------
The Wilson coefficients 
${\tilde{L}}_{2,q}^{\sf PS}\left(n_f,{Q^2}/{m^2},{m^2}/{\mu^2} \right)$
and ${\tilde{L}}_{2,g}^{\sf S}\left(n_f,{Q^2}/{m^2},{m^2}/{\mu^2} \right)$ 
have been calculated completely for general values of $N$ in \cite{ABKSW11NPB}. 

The renormalization group implies the following representation for the set
of $(n_f +1)$ (massless) parton densities expressed in terms of $n_f$ 
parton densities \cite{BMSN98}~:
%-------------------------------------------------------------------------------
\begin{eqnarray}
\label{HPDF1}
f_k(n_f+1,\mu^2,m^2,N) + f_{\overline{k}}(n_f+1,\mu^2,m^2,N)
&=& A_{qq,Q}^{\rm NS}\left(n_f,\frac{\mu^2}{m^2},N\right)
\cdot \bigl[f_k(n_f,\mu^2,N)
\nonumber\\
 && \hspace*{3.8cm}
+ f_{\overline{k}}(n_f,\mu^2,N)\bigr]
\nonumber\\
& & \hspace*{-4mm} + \tilde{A}_{qq,Q}^{\rm
PS}\left(n_f,\frac{\mu^2}{m^2},N\right)
\cdot \Sigma(n_f,\mu^2,N)
\nonumber\\
& & \hspace*{-4mm} + \tilde{A}_{qg,Q}\left(n_f,\frac{\mu^2}{m^2},N\right)
\cdot G(n_f,\mu^2,N),
\\
%----
\label{fQQB}
f_Q(n_f+1,\mu^2,m^2,N) + f_{\overline{Q}}(n_f+1,\mu^2,m^2,N)
%\nonumber\\
&=&
{A}_{Qq}^{\rm PS}\left(n_f,\frac{\mu^2}{m^2},N\right)
\cdot \Sigma(n_f,\mu^2,N)   
\nonumber\\ && \hspace*{-4mm} 
+ {A}_{Qg}\left(n_f,\frac{\mu^2}{m^2},N\right)
\cdot G(n_f,\mu^2,N)~.
%----
\end{eqnarray}
%-------------------------------------------------------------------------------
Here $f_Q (f_{\bar{Q}})$ are the heavy quark densities.
The flavor singlet, non--singlet and gluon densities for $(n_f+1)$ flavors are
given by
%-------------------------------------------------------------------------------
%-------------------------------------------------------------------------------
\begin{eqnarray}
\Sigma(n_f+1,\mu^2,m^2,N)
&=& \Biggl[A_{qq,Q}^{\rm NS}\left(n_f, \frac{\mu^2}{m^2},N\right) +
          n_f \tilde{A}_{qq,Q}^{\rm PS}\left(n_f, \frac{\mu^2}{m^2},N\right)
  \nonumber \\ &&
         + {A}_{Qq}^{\rm PS}\left(n_f, \frac{\mu^2}{m^2},N\right)
        \Biggr]
\cdot \Sigma(n_f,\mu^2,N) \nonumber
\\   
%\end{eqnarray} \begin{eqnarray}
& & \hspace*{-3mm} + \left[n_f \tilde{A}_{qg,Q}\left(n_f,
\frac{\mu^2}{m^2},N\right) +
          {A}_{Qg}\left(n_f, \frac{\mu^2}{m^2},N\right)
\right] 
\cdot G(n_f,\mu^2,N)
\nonumber\\
\\
\Delta(n_f+1,\mu^2,m^2,N) &=&
  f_k(n_f+1,\mu^2,N)
+ f_{\overline{k}}(n_f+1,\mu^2,m^2,N)
\nonumber\\ &&
- \frac{1}{n_f+1}
\Sigma(n_f+1,\mu^2,m^2,N)\\
\label{HPDF2}  
G(n_f+1,\mu^2,m^2,N) &=& A_{gq,Q}\left(n_f,
\frac{\mu^2}{m^2},N\right)
                    \cdot \Sigma(n_f,\mu^2,N)
\nonumber\\
&&
+ A_{gg,Q}\left(n_f, \frac{\mu^2}{m^2},N\right)
                    \cdot G(n_f,\mu^2,N)~.
\end{eqnarray}
%-------------------------------------------------------------------------------
Any relation between the $(n_f +1)$- and $n_f$-parton density can only 
contain {\sf universal}, i.e. process-independent, quantities. 

Note that the {\sf new} parton densities depend on the renormalized heavy
quark mass $m^2$. As outlined above, the corresponding relations for the
operator matrix elements depend on the mass--renormalization scheme, with
$m = m(a_s(\mu^2))$ in the $\overline{\rm MS}$ scheme, which we will apply below.
These equations describe the transition of one heavy quark becoming light at the time
referring to the scale $\mu^2$.

The matching scales $\mu^2$ are often chosen as $\mu^2 = m^2$. The comparison of
the results in complete calculations to those in which flavor thresholds are 
matched in the VFNS allows in principle to determine the relevant matching scale.
In an analysis of the various deep-inelastic structure function sum rules 
\cite{Blumlein:1998sh} it has been shown  that the scale $\mu^2$ turns out to be 
significantly different of $m^2$. This is not unexpected since mass effects do 
not turn into the behaviour of the massless case close to the production threshold.

The resummation of large logs, as being performed in the VFNS, has to be performed
at very high scales. As has been shown in \cite{GRS94NPB} this is not the case in
the kinematic range at HERA. A smooth transition from the threshold region to 
asymptotic scales has been proposed in terms of the BMSN-scheme \cite{BMSN98},
%--------------------------------------------------------------------
{\small
\begin{eqnarray}
F_2^{c\bar{c}}(x,Q^2,n_f = 4) = F_2^{c\bar{c}, \rm FFNS}(x,Q^2,n_f=3) +
                                F_2^{c\bar{c}, \rm asymp}(x,Q^2,n_f=4)
                                -F_2^{c\bar{c}, \rm asymp}(x,Q^2,n_f=3)~,
\end{eqnarray}}
%--------------------------------------------------------------------

\noindent
which is found to be in excellent agreement with the HERA data \cite{Alekhin:2009ni}.
There is a series of other proposals to match between the threshold and asymtotic
region \cite{ACOT,Forte:2010ta,Thorne:2006qt}, partly with a faster transition to
the massless case. Here precise data on $F_2^{c\bar{c}}(x,Q^2)$ are helpful to 
distinguish between different descriptions. We would like to mention that a correct 
treatment of the heavy flavor corrections is of instrumental importance in the
QCD analysis of the complete structure functions $F_2(x,Q^2)$, which has been measured
to a precision of $O(1\%)$ \cite{herapdf:2009wt}.

%%%%%%%%%%%%%%%%%%%%%%%%%%%%%%%%%%%%%%%%%%%%%%%%%%%%%%%%%%%%%%%%%%%%%%%%
%           Calculation and Results
%%%%%%%%%%%%%%%%%%%%%%%%%%%%%%%%%%%%%%%%%%%%%%%%%%%%%%%%%%%%%%%%%%%%%%%
\section{The \boldmath{$O(\alpha_s^3n_fT_F^2)$} contributions to \boldmath{$A_{gg,Q}$} 
and \boldmath{$A_{gq,Q}$}}
\label{sec:3}
%%%%%%%%%%%%%%%%%%%%%%%%%%%%%%%%%%%%%%%%%%%%%%%%%%%%%%%%%%%%%%%%%%%%%%%

\vspace{1mm}
\noindent
The OMEs $A_{gq,Q}$ and $A_{gg,Q}$ are expectation values $\langle j|O_g|j\rangle$, $i,j=q,g$ 
of the gluonic operator  
%--------------------------------------------------------------------
\begin{eqnarray}
  O_{g,\mu_1,...,\mu_N}
  = 2 i^{N-2} \mathbf{S}\text{Sp}
  [F_{\mu_1\alpha} D_{\mu_2} ... D_{\mu_N-1}F_{\mu_N}^{\alpha}]
  - \text{trace terms}
  ~.
\end{eqnarray}
%--------------------------------------------------------------------
between massless on-shell external states.
The corresponding massive OMEs $A_{gq,Q},A_{gg,Q}$ were calculated to
$O(\alpha_s^2)$ in \cite{BMSN98} and including also terms linear in $\ep$ in 
\cite{BBK09PLB} correcting the previous result.

The renormalized expressions $A_{gq,Q}$ and $A_{gg,Q}$ to $O(a_s^3)$ were derived in 
\cite{BBK09NPB}.
In the $\overline{\rm MS}$ scheme with the heavy quark mass $m$ on-shell they are given by~:
%--------------------------------------------------------------------
\begin{eqnarray}
\label{eqAgqQ}
    A_{gq,Q}^{(3), \MS}&=&
                     -\frac{\gamma_{gq}^{(0)}}{24}
                      \Biggl\{
                          \gamma_{gq}^{(0)}\hat{\gamma}_{qg}^{(0)}
                         +\Bigl(
                              \gamma_{qq}^{(0)}
                             -\gamma_{gg}^{(0)}
                             +10\beta_0
                             +24\beta_{0,Q}
                                     \Bigr)\beta_{0,Q}
                      \Biggr\}
                           \ln^3 \Bigl(\frac{m^2}{\mu^2}\Bigr)
                     +\frac{1}{8}\Biggl\{
                         6\gamma_{gq}^{(1)}\beta_{0,Q}
\N\\ &&
                        +\hat{\gamma}_{gq}^{(1)}\Bigl(
                                      \gamma_{gg}^{(0)}
                                     -\gamma_{qq}^{(0)}
                                     -4\beta_0
                                     -6\beta_{0,Q}
                                                 \Bigr)
                        +\gamma_{gq}^{(0)}\Bigl(
                                       \hat{\gamma}_{qq}^{(1), {\sf
NS}}
                                      +\hat{\gamma}_{qq}^{(1), {\sf
PS}}
                                      -\hat{\gamma}_{gg}^{(1)}
                                      +2\beta_{1,Q}
                                                 \Bigr)
                      \Biggr\}
                           \ln^2 \Bigl(\frac{m^2}{\mu^2}\Bigr)
\N\\ &&
                     +\frac{1}{8}\Biggl\{
                              4\hat{\gamma}_{gq}^{(2)}
                            + 4a_{gq,Q}^{(2)}         \Bigl(
                                    \gamma_{gg}^{(0)}
                                   -\gamma_{qq}^{(0)}
                                   -4\beta_0
                                   -6\beta_{0,Q}
                                                       \Bigr)
                            + 4\gamma_{gq}^{(0)}       \Bigl(
                                      a_{qq,Q}^{(2),{\sf NS}}
                                     +a_{Qq}^{(2),{\sf PS}}
                                     -a_{gg,Q}^{(2)}
\N\\ &&
                                     +\beta_{1,Q}^{(1)}
                                                       \Bigr)
                            + \gamma_{gq}^{(0)}\zeta_2 \Bigl(
                               \gamma_{gq}^{(0)}\hat{\gamma}_{qg}^{(0)}
                               +\Bigl[
                                        \gamma_{qq}^{(0)}
                                       -\gamma_{gg}^{(0)}
                                       +12\beta_{0,Q}
                                       +10\beta_0
                                            \Bigr]\beta_{0,Q}
                                                       \Bigr)
                      \Biggr\}
                           \ln \Bigl(\frac{m^2}{\mu^2}\Bigr)
\N\\ &&
                  + \overline{a}_{gq,Q}^{(2)} \Bigl(
                                       \gamma_{qq}^{(0)}
                                      -\gamma_{gg}^{(0)}
                                      +4\beta_0
                                      +6\beta_{0,Q}
                                             \Bigr)
                  + \gamma_{gq}^{(0)} \Bigl(
                                       \overline{a}_{gg,Q}^{(2)}
                                      -\overline{a}_{Qq}^{(2),{\sf
PS}}
                                      -\overline{a}_{qq,Q}^{(2),{\sf
NS}}
                                             \Bigr)
                -\gamma_{gq}^{(0)}\beta_{1,Q}^{(2)}
\N\\ && 
                -\frac{\gamma_{gq}^{(0)}\zeta_3}{24} \Bigl(
                           \gamma_{gq}^{(0)}\hat{\gamma}_{qg}^{(0)}
                          +\Bigl[
                                   \gamma_{qq}^{(0)}
                                  -\gamma_{gg}^{(0)}
                                  +10\beta_0
                                      \Bigr]\beta_{0,Q}
                                             \Bigr)
                -\frac{3\gamma_{gq}^{(1)}\beta_{0,Q}\zeta_2}{8}
                +2 \delta m_1^{(-1)} a_{gq,Q}^{(2)}
\N\\ &&
                +\delta m_1^{(0)} \hat{\gamma}_{gq}^{(1)}
                +4 \delta m_1^{(1)} \beta_{0,Q} \gamma_{gq}^{(0)}
                +a_{gq,Q}^{(3)},
\end{eqnarray}
\begin{eqnarray}
\label{eqAggQ}
  A_{gg,Q}^{(3), \MS}&=&
                    \frac{1}{48}\Biggl\{
                            \gamma_{gq}^{(0)}\hat{\gamma}_{qg}^{(0)}
                                \Bigl(
                                        \gamma_{qq}^{(0)}
                                       -\gamma_{gg}^{(0)}
                                       -6\beta_0
                                       -4n_f\beta_{0,Q}
                                       -10\beta_{0,Q}
                                \Bigr)
                           -4
                                \Bigl(
                                        \gamma_{gg}^{(0)}\Bigl[
                                            2\beta_0
                                           +7\beta_{0,Q}
                                                         \Bigr]
\N
\\ 
&&
                                       +4\beta_0^2
                                       +14\beta_{0,Q}\beta_0
                                       +12\beta_{0,Q}^2
                                \Bigr)\beta_{0,Q}
                     \Biggr\}
                     \ln^3 \Bigl(\frac{m^2}{\mu^2}\Bigr)
                    +\frac{1}{8}\Biggl\{
                            \hat{\gamma}_{qg}^{(0)}
                                \Bigl(
                                        \gamma_{gq}^{(1)}
                                       +(1-n_f)\hat{\gamma}_{gq}^{(1)}
                                \Bigr)
\N\\ &&
                           +\gamma_{gq}^{(0)}\hat{\gamma}_{qg}^{(1)}
                           +4\gamma_{gg}^{(1)}\beta_{0,Q}
                           -4\hat{\gamma}_{gg}^{(1)}[\beta_0+2\beta_{0,Q}]
                           +4[\beta_1+\beta_{1,Q}]\beta_{0,Q}
\N\\ &&
                           +2\gamma_{gg}^{(0)}\beta_{1,Q}
                     \Biggr\}
                     \ln^2 \Bigl(\frac{m^2}{\mu^2}\Bigr)
                    +\frac{1}{16}\Biggl\{
                            8\hat{\gamma}_{gg}^{(2)}
                           -8n_fa_{gq,Q}^{(2)}\hat{\gamma}_{qg}^{(0)}
                           -16a_{gg,Q}^{(2)}(2\beta_0+3\beta_{0,Q})
\N\\ &&
                           +8\gamma_{gq}^{(0)}a_{Qg}^{(2)}
                           +8\gamma_{gg}^{(0)}\beta_{1,Q}^{(1)}
                   +\gamma_{gq}^{(0)}\hat{\gamma}_{qg}^{(0)}\zeta_2
                                \Bigl(
                                        \gamma_{gg}^{(0)}
                                       -\gamma_{qq}^{(0)}
                                       +6\beta_0
                                       +4n_f\beta_{0,Q}
                                       +6\beta_{0,Q}
                                \Bigr)
\N\\ &&
                   +4\beta_{0,Q}\zeta_2
                                \Bigl( 
                                       \gamma_{gg}^{(0)}
                                      +2\beta_0
                                \Bigr)
                                \Bigl(
                                       2\beta_0
                                      +3\beta_{0,Q}
                                \Bigr)
                     \Biggr\}
                     \ln \Bigl(\frac{m^2}{\mu^2}\Bigr)
                   +2(2\beta_0+3\beta_{0,Q})\overline{a}_{gg,Q}^{(2)}\N
\N\\ &&
                   +n_f\hat{\gamma}_{qg}^{(0)}\overline{a}_{gq,Q}^{(2)}
                   -\gamma_{gq}^{(0)}\overline{a}_{Qg}^{(2)}
                   -\beta_{1,Q}^{(2)} \gamma_{gg}^{(0)}
                   +\frac{\gamma_{gq}^{(0)}\hat{\gamma}_{qg}^{(0)}\zeta_3}{48}
                                \Bigl(
                                        \gamma_{qq}^{(0)}
                                       -\gamma_{gg}^{(0)}
                                       -2[2n_f+1]\beta_{0,Q}
\N\\ &&
                                       -6\beta_0
                                \Bigr)
                   +\frac{\beta_{0,Q}\zeta_3}{12}
                                \Bigl(
                                        [\beta_{0,Q}-2\beta_0]\gamma_{gg}^{(0)}
                                       +2[\beta_0+6\beta_{0,Q}]\beta_{0,Q}
                                       -4\beta_0^2
                                \Bigr)
\N\\ &&
                   -\frac{\hat{\gamma}_{qg}^{(0)}\zeta_2}{16}
                                \Bigl(
                                        \gamma_{gq}^{(1)}
                                       +\hat{\gamma}_{gq}^{(1)}
                                \Bigr)
                   +\frac{\beta_{0,Q}\zeta_2}{8}
                                \Bigl(
                                        \hat{\gamma}_{gg}^{(1)}
                                      -2\gamma_{gg}^{(1)}
                                      -2\beta_1
                                      -2\beta_{1,Q}
                                \Bigr)
                           +\frac{\delta m_1^{(-1)}}{4}
                                \Bigl(
                                     8 a_{gg,Q}^{(2)}
\N\\ &&
                                    +24 \delta m_1^{(0)} \beta_{0,Q}
                                    +8 \delta m_1^{(1)} \beta_{0,Q} 
                                    +\zeta_2 \beta_{0,Q} \beta_0
                                    +9 \zeta_2 \beta_{0,Q}^2
                                \Bigr)
                           +\delta m_1^{(0)}
                                \Bigl(
                                     \beta_{0,Q} \delta m_1^{(0)}
                                    +\hat{\gamma}_{gg}^{(1)}
                                \Bigr)
\N\\ &&
                           +\delta m_1^{(1)}
                                \Bigl(
                                     \hat{\gamma}_{qg}^{(0)} \gamma_{gq}^{(0)}
                                    +2 \beta_{0,Q} \gamma_{gg}^{(0)}
                                    +4 \beta_{0,Q} \beta_0
                                    +8 \beta_{0,Q}^2
                                \Bigr)
                           -2 \delta m_2^{(0)} \beta_{0,Q}
                 +a_{gg,Q}^{(3)}~. 
        \label{Agg3QMSren}
   \end{eqnarray}
%--------------------------------------------------------------------
Here $\delta m_i^{(k)}$ are expansion coefficients of the unrenormalized mass,
$\beta_i,\beta_{i,Q}$ are coefficients of the $\beta$-functions (including mass 
effects), $\zeta_k$ is the Riemann--$\zeta$ function with $k \in \mathbb{N} \backslash 
\{0,1\}$, $a^{(2)}_{ij}, \overline{a}^{(2)}_{ij}$
are two loop contributions to order $\ep^0$ and $\ep^1$ respectively,
and $\gamma_{ij},\hat{\gamma}_{ij}$ are the anomalous dimensions, and quantities with
a hat or a tilde are defined by
%--------------------------------------------------------------------
\begin{eqnarray}
  \hat{f} = f(n_f+1) - f(n_f), \hspace{1.3cm} \tilde{f} = \frac{1}{n_f}~f,
\end{eqnarray}
%--------------------------------------------------------------------
see Ref.~\cite{BBK09NPB}.
%--------------------------------------------------------------------
The unreormalized OME $\Ahathat_{gg,Q}^{(3)}$ also receives contributions from the 
vacuum polarization insertions on the external lines
%--------------------------------------------------------------------
\begin{eqnarray}
\hat{\Pi}^{ab}_{\mu\nu}(p^2,\hat{m}^2,\mu^2,\hat{a}_s^2) 
 &=& i\delta^{ab}\left[-g_{\mu\nu} p^2 + p_\mu p_\nu\right]
  \sum_{k-1}^{\infty}\hat{a}_s^k \hat{\Pi}^{(k)}(p^2,\hat{m}^2,\mu^2)
\\
  \hat{\Pi}^{(k)} &\equiv& \hat{\Pi}^{(k)}(0,\hat{m}^2,\mu^2)
\end{eqnarray}
%--------------------------------------------------------------------
such that
%--------------------------------------------------------------------
\begin{eqnarray}
   \Ahathat_{gg,Q}^{(3)}
   &=&
            \Ahathat_{gg,Q}^{(3), \text{1PI}}
           -\hat{\Pi}^{(3)}
           -\Ahathat_{gg,Q}^{(2), \text{1PI}}
            \hat{\Pi}^{(1)}
           -2\Ahathat_{gg,Q}^{(1)}
            \hat{\Pi}^{(2)}
           +\Ahathat_{gg,Q}^{(1)}
            \hat{\Pi}^{(1)}
            \hat{\Pi}^{(1)}
\\
  &\equiv&
   \frac{a_{gg,Q}^{(3,0)}}{\ep^3}
  +\frac{a_{gg,Q}^{(3,1)}}{\ep^2}
  +\frac{a_{gg,Q}^{(3,2)}}{\ep}
  +a_{gg,Q}^{(3)}~.
\end{eqnarray}
%--------------------------------------------------------------------
All contributions to (\ref{eqAgqQ},\ref{eqAggQ}) but the constant terms 
$a_{ij,Q}^{(3)}$ are 
known~\cite{BMSN96,BBK07NPB,BBKS08NPB,BMSN98,BBK09PLB,Vogt:2004mw}. In particular, all the
logarithmic contributions have already been obtained for general values of the Mellin variable $N$,
\cite{Bierenbaum:2010jp}.

In the following we calculate the contributions $O(a_s^3 n_f T_F^2 C_{F,A})$ to the 
massive gluonic OMEs. The Feynman diagrams are generated by {\tt QGRAF} 
\cite{Nogueira:1991ex} and the extension allowing to include local operators 
\cite{BBK09NPB}. The color-algebra is performed using \cite{vanRitbergen:1998pn}.
For a large part of the calculation we use {\tt FORM} \cite{FORM}. The momentum 
integrals are performed introducing a Feynman parameterization. The Feynman parameter 
integrals are then rewritten in terms of hypergeometric functions (${}_2F_1$,${}_3F_2$), 
which are represented in terms of absolutely convergent series. The 
resulting sums, which 
may still contain finite sums due to binomial expansions, are then processed applying 
the symbolic summation technology, which is encoded in the package {\tt Sigma} 
\cite{SIGMA}
and making use of a large number of algorithms for processing multi sums using the 
package {\tt EvaluateMultiSums} \cite{EvalMulSum,Blumlein:2012hg}.  Additionally it is very useful to 
reduce such sums to a smaller number of `key sums', by synchronization of the summation 
ranges and algebraic reduction of the summands. This step helped to reduce the size of 
the terms from 2GByte to 7.6MByte and the number of sums from 2419 to 29.  The 
algorithms for this step are implemented in the package {\tt SumProduction} 
\cite{SumProduction}. Details of the corresponding technique are described in 
\cite{EvalMulSum,Blumlein:2012hg}. The corresponding expressions have simplified
using mutual relations and methods applicable to the respective classes of sums encoded in 
the package {\tt HarmonicSums} \cite{Harmonicsums}. The results for the individual diagrams
have been checked comparing to the moments obtained in \cite{BBK09NPB} using the code {\tt MATAD}
\cite{Steinhauser:2000ry}. 
The constant contributions $a_{gj,Q}^{(3)},~j = q,g$ to (\ref{eqAgqQ},\ref{eqAggQ}) 
read~:

\vspace*{-3mm}
%--------------------------------------------------------------------
\begin{eqnarray}
  a_{gq,Q}^{(3),n_fT_F^2}
  &=& C_F T_F^2 n_f \Biggl\{ 
  -\frac{16 \left(N^2+N+2\right)}{9  (N-1) N (N+1)}
   \left( \frac{1}{3} S_{1}^3 + S_{2} S_{1} + \frac{2}{3} S_{3} +
         14 \zeta_3 + 3 S_{1} \zeta_2 \right)
\N\\ & &
  +\frac{16 \left(8 N^3+13 N^2+27 N+16\right)}{27 (N-1) N (N+1)^2} 
   \left( 3 \zeta_2 + S_{1}^2 + S_{2}\right)
\N\\ & &
  -\frac{32 \left(35 N^4+97 N^3+178 N^2+180 N+70\right)}{27 (N-1) N
(N+1)^3} S_{1}
\N\\ & &
  +\frac{32 \left(1138 N^5+4237 N^4+8861 N^3+11668 N^2+8236
N+2276\right)}{243 (N-1) N (N+1)^4}
 \Biggr\}~.
\\
%\end{eqnarray}
%--------------------------------------------------------------------
%--------------------------------------------------------------------
%\begin{eqnarray}
  a_{gg,Q}^{(3),n_f T_F^2}
  &=& n_f T_F^2 \Biggl\{
  C_A 
  \frac{1}{(N-1) (N+2)}
  \Biggl[
    \frac{4 P_{1}}{27 N^2 (N+1)^2} S_{1}^2
   +\frac{8 P_{2}}{729 N^3 (N+1)^3} S_{1}
\N\\
&&   +\frac{160}{27} (N-1) (N+2) \zeta_2 S_{1}
   -\frac{448}{27} (N-1) (N+2) \zeta_3 S_{1}
   +\frac{P_{3}}{729 N^4 (N+1)^4}
\N\\
&&
   -\frac{2 P_{4}}{27 N^2 (N+1)^2} \zeta_2
   +\frac{56 \left(3 N^4+6 N^3+13 N^2+10 N+16\right)}{27 N (N+1)} \zeta_3
   -\frac{4 P_{5}}{27 N^2 (N+1)^2} S_{2}
  \Biggr]
\N\\&&
  +C_F 
   \frac{1}{(N-1) (N+2)}
   \Biggl[
     \frac{112 \left(N^2+N+2\right)^2}{27 N^2 (N+1)^2} S_{1}^3
    -\frac{16 P_{6}}{27 N^3 (N+1)^3} S_{1}^2
\N\\&&
    +\frac{32 P_{7}}{81 N^4 (N+1)^4} S_{1}
    +\frac{16 \left(N^2+N+2\right)^2}{3 N^2 (N+1)^2} \zeta_2 S_{1}
    +\frac{16 \left(N^2+N+2\right)^2}{3 N^2 (N+1)^2} S_{2} S_{1}
\N\\&&
    -\frac{32 P_{8}}{243 N^5 (N+1)^5}
    -\frac{16 P_{9}}{9 N^3 (N+1)^3} \zeta_2
    +\frac{448 \left(N^2+N+2\right)^2}{9 N^2 (N+1)^2} \zeta_3
    +\frac{16 P_{10}}{9 N^3 (N+1)^3} S_{2}
\N\\&&
    -\frac{160 \left(N^2+N+2\right)^2}{27 N^2 (N+1)^2} S_{3}
   \Biggr]\Biggr\}~,
\end{eqnarray}
%--------------------------------------------------------------------
where the polynomials $P_i$ are given by
%--------------------------------------------------------------------
\begin{eqnarray}
P_{1} &=& 16 N^5+41 N^4+2 N^3+47 N^2+70 N+32 \\
P_{2} &=& 6944 N^8+26480 N^7+23321 N^6-15103 N^5-39319 N^4-27001 N^3-11178 N^2 \N\\&&
-2016 N+864 \\
P_{3} &=& 4809 N^{10}+24045 N^9-182720 N^8-854414 N^7-1522031 N^6-1472927 N^5
\N\\&&
-758234 N^4 
-126080 N^3-1152 N^2-50688 N-24192 \\
P_{4} &=& 3 N^6+9 N^5+307 N^4+599 N^3+746 N^2+448 N+96 \\
P_{5} &=& 40 N^6+112 N^5-3 N^4-166 N^3-301 N^2-210 N-96 \\
P_{6} &=& 44 N^6+123 N^5+386 N^4+543 N^3+520 N^2+248 N+24 \\
P_{7} &=& 205 N^8+856 N^7+3169 N^6+6484 N^5+7310 N^4+4722 N^3+1534 N^2
\N\\ &&
+48 N-72 \\
P_{8} &=& 1976 N^{10}+9385 N^9+24088 N^8+38989 N^7+50214 N^6+53872 N^5+35219 N^4 \N\\&&
+6890 N^3-4233 N^2-2844 N-756 \\
P_{9} &=& 14 N^6+33 N^5+59 N^4+39 N^3+55 N^2+20 N-12\\
P_{10} &=& 4 N^6+3 N^5-50 N^4-129 N^3-100 N^2-56 N-24~.
\end{eqnarray}
%--------------------------------------------------------------------
Here $S_{b,\vec{a}} = \equiv S_{b,\vec{a}}(N) = \sum_{n=1}^N {\rm sign(b)}^n S_{\vec{a}}(N)/n^{|b|};~ 
S_{\emptyset} = 1$ denote the harmonic sums \cite{HSUM}
which only occur as single harmonic sums in the present calculation.

It is convenient to express the renormalized OMEs $A_{gj,Q},~j=q,g$ also referring
to the heavy quark mass in the $\overline{\rm MS}$ scheme, cf.~\cite{BBK09NPB}.
The OMEs $A_{gq,Q}^{(3),n_fT_F^2}$ and $A_{gg,Q}^{(3),n_fT_F^2}$ read~:
%--------------------------------------------------------------------
\begin{eqnarray}
\label{eq:OME1}
  A_{gq,Q,C_FT_F^2n_f}^{(3), \MS}
  &=&
  C_F n_f T_F^2 \Biggl\{\frac{32 \left(N^2+N+2\right)}{9 (N-1) N (N+1)}
  \ln^3\left(\frac{\bar{m}^2}{\mu^2}\right)
\N\\&&
 +\Biggl[
   -\frac{16 \left(N^2+N+2\right)}{3 (N-1) N (N+1)}
    \left( S_{1}^2 + S_{2}\right)
   +\frac{32 \left(8 N^3+13 N^2+27 N+16\right)}{9 (N-1) N (N+1)^2}
S_{1}
\N\\&&
   +\frac{32 \left(19 N^4+81 N^3+86 N^2+80 N+38\right)}{27 (N-1) N
(N+1)^3}
  \Biggr] \ln\left(\frac{\bar{m}^2}{\mu^2}\right)
\N\\
&&
 +\Biggl[
    \frac{32 \left(N^2+N+2\right)}{27 (N-1) N (N+1)}
    \left( S_{1}^3 + 3 S_{2} S_{1} + 2 S_{3} - 24 \zeta_3\right)
\N\\ &&
   -\frac{32 \left(8 N^3+13 N^2+27 N+16\right)}{27 (N-1) N (N+1)^2}
    \left( S_{1}^2 + S_{2}\right)
\N\\&&
   +\frac{64 \left(4 N^4+4 N^3+23 N^2+25 N+8\right)}{27 (N-1) N
(N+1)^3} S_{1}
\N\\&&
   +\frac{64 \left(197 N^5+824 N^4+1540 N^3+1961 N^2+1388
N+394\right)}{243 (N-1) N (N+1)^4}
  \Biggr]\Biggr\}
\\
%----------------------------------------------------------------------
\label{eq:OME2}
 A_{gg,Q}^{(3),n_fT_F^2,\MS}
 &=& 
  n_f T_F^2 
  \Biggl\{
  \Biggl(
    C_F \frac{64 \left(N^2+N+2\right)^2}{9 (N-1) N^2 (N+1)^2 (N+2)}
\N\\
& &
    +C_A \Biggl[
      \frac{128 \left(N^2+N+1\right)}{27 (N-1) N (N+1) (N+2)}
     -\frac{64}{27} S_{1}
    \Biggr]
  \Biggr) \ln^3\left(\frac{\bar{m}^2}{\mu^2}\right)
\N
\end{eqnarray}
\begin{eqnarray}
\\ & &
  -C_F \frac{16}{3} \ln^2\left(\frac{\bar{m}^2}{\mu^2}\right)
  +\Biggl(
    C_A 
    \frac{1}{(N-1) (N+2)}
    \Biggl[
      -\frac{4 P_{11}}{81 N^3 (N+1)^3}
\N\\
\\ 
& &
      -\frac{16 P_{12}}{81 N^2 (N+1)^2} S_{1}
    \Biggr]
   +C_F 
    \frac{1}{(N-1) (N+2)}
    \Biggl[
       \frac{16 \left(N^2+N+2\right)^2}{N^2 (N+1)^2}
       \left( S_{1}^2 - \frac{5}{3} S_{2} \right)
\N\\ & &
      -\frac{4 P_{13}}{9 N^4 (N+1)^4}
      -\frac{32 P_{14}}{3 N^3 (N+1)^3} S_{1}
    \Biggr]
  \Biggr) \ln\left(\frac{\bar{m}^2}{\mu^2}\right)
\N\\
& &
  +C_A 
  \frac{1}{(N-1) (N+2)}
  \Biggl[
    -\frac{4 P_{15}}{27 N^2 (N+1)^2} S_{1}^2
    -\frac{8 P_{16}}{729 N^3 (N+1)^3} S_{1}
\N\\ & &
    +\frac{512}{27} (N-1) (N+2) \zeta_3 S_{1}
    -\frac{2 P_{17}}{729 N^4 (N+1)^4}
    -\frac{1024 \left(N^2+N+1\right)}{27 N (N+1)} \zeta_3
\N\\ & &
    +\frac{4 P_{18}}{27 N^2 (N+1)^2} S_{2}
  \Biggr]
\N\\ & &
  +C_F 
  \frac{1}{(N-1) (N+2)}
  \Biggl[
     \frac{64 \left(N^2+N+2\right)^2}{9 N^2 (N+1)^2}
     \left(-\frac{1}{3} S_{1}^3 - 8 \zeta_3 + \frac{4}{3} S_{3}
\right)
\N\\ & &
    +\frac{32 P_{19}}{27 N^3 (N+1)^3} S_{1}^2
    -\frac{64 P_{20}}{81 N^4 (N+1)^4} S_{1}
    -\frac{32 P_{21}}{243 N^5 (N+1)^5}
\N\\ & &
    -\frac{32 P_{22}}{3 N^3 (N+1)^3} S_{2}
  \Biggr]
  \Biggr\}~,
\end{eqnarray}
%--------------------------------------------------------------------
with the polynomials
%--------------------------------------------------------------------
\begin{eqnarray}
P_{11} &=& 297 N^8+1188 N^7+640 N^6-2094 N^5-1193 N^4+2874 N^3+5008
N^2
\N\\&&+3360 N+864 \\
P_{12} &=& 136 N^6+390 N^5+19 N^4-552 N^3-947 N^2-630 N-288 \\
P_{13} &=& 15 N^{10}+75 N^9-48 N^8-866 N^7-2985 N^6-6305 N^5-8206
N^4-7656 N^3 \N\\
&&
         -4648 N^2-1600 N-288 \\
P_{14} &=& 5 N^5+52 N^4+109 N^3+90 N^2+48 N+16 \\
%\end{eqnarray}
%\begin{eqnarray}
P_{15} &=& 4 N^5+17 N^4+14 N^3+71 N^2+70 N+32 \\
P_{16} &=& 3008 N^8+11600 N^7+9197 N^6-10255 N^5-27739 N^4-24745
N^3-12474 N^2\N\\&&
         -2016 N+864 \\
P_{17} &=& 4185 N^{10}+20925 N^9+1892 N^8-117118 N^7-222151 N^6-176863
N^5-41446 N^4 \N\\&&
         +22304 N^3-1296 N^2-18432 N-6912 \\
P_{18} &=& 16 N^6+52 N^5-3 N^4-106 N^3-277 N^2-210 N-96 \\
P_{19} &=& 10 N^6+30 N^5+109 N^4+168 N^3+155 N^2+76 N+12 \\
P_{20}&=& 38 N^8+206 N^7+962 N^6+2246 N^5+2509 N^4+1542 N^3+509 N^2+24
N-36 \\
P_{21}&=& 123 N^{12}+738 N^{11}+691 N^{10}-3526 N^9-14521 N^8-29458
N^7-39189 N^6
\N\\&&
-37672 N^5 
         -21920 N^4-3914 N^3+2856 N^2+1872 N+432 \\
P_{22}&=& 2 N^6+4 N^5+N^4-10 N^3-5 N^2-4 N-4~.
\end{eqnarray}
%--------------------------------------------------------------------

As has been noted before \cite{BBK09NPB}, the above results are free of $\zeta_2$, 
which is common to all massive OMEs, and hence is a particular feature of representing 
also the mass in the {\MSbar}~scheme.  Furthermore we note, that the 
$\ln^2\left({\bar{m}^2}/{\mu^2}\right)$-contribution to $A_{gg,Q,C_FT_F^2n_f}^{(3), \MS}$ 
is particularly simple, while the corresponding contribution to 
$A_{gq,Q,C_FT_F^2n_f}^{(3), \MS}$ vanishes.  

As a by-product of the calculation we obtain the corresponding contributions to the
anomalous dimensions from the single pole term $1/\ep$ resp.\ the linear logarithmic 
contribution, cf.~(\ref{eqAgqQ},\ref{eqAggQ}),
%--------------------------------------------------------------------
\begin{eqnarray}
\label{eq:ggq3}
 \hat{\gamma}_{gq}^{(2),n_f}
 &=&
 n_f T_F^2 C_F \Biggl(
    -\frac{64 \left(N^2+N+2\right)}{3 (N-1) N (N+1)}
   \left(  S_1^2 + S_2\right)
   +\frac{128 \left(8 N^3+13 N^2+27 N+16\right)}{9 (N-1) N (N+1)^2} S_1
\nonumber\\&&
   -\frac{128 \left(4 N^4+4 N^3+23 N^2+25 N+8\right)}{9 (N-1) N (N+1)^3}
 \Biggr)
 ~,
\\
%--------------------------------------------------------------------
\label{eq:ggg3}
  \hat{\gamma}_{gg}^{(2),n_f} &=&
  n_f T_F^2 C_A \Biggl[
    -\frac{32 P_{23}}{27 (N-1) N^2 (N+1)^2 (N+2)} S_{1}
%\N\\&&
    -\frac{8 P_{24}
}{27 (N-1) N^3 (N+1)^3 (N+2)}
  \Biggr]
\N\\&&
 +n_f T_F^2 C_F \Biggl[
     \frac{64 \left(N^2+N+2\right)^2}{3 (N-1) N^2 (N+1)^2 (N+2)}
     \left( S_{1}^2 - 3 S_{2} \right) 
\N\\&&
    +\frac{128 P_{25}}
{9 (N-1) N^3 (N+1)^3 (N+2)} S_{1}
    -\frac{16 P_{26}}{27 (N-1) N^4 (N+1)^4 (N+2)}
 \Biggr]
 ~,
\end{eqnarray}
%--------------------------------------------------------------------
where
%--------------------------------------------------------------------
\begin{eqnarray}
P_{23} &=& 8 N^6+24 N^5-19 N^4-78 N^3-253 N^2-210 N-96
\\
P_{24} &=& 87 N^8+348 N^7+848 N^6+1326 N^5+2609 N^4+3414 N^3+2632 N^2+1088 N
\N\\ &&
+192
\\
P_{25} &=& 4 N^6+3 N^5-50 N^4-129 N^3-100 N^2-56 N-24~,
\\
P_{26} &=& 33 N^{10}+165 N^9+256 N^8-542 N^7-3287 N^6-8783 N^5-11074 N^4-9624 N^3 
\N\\ &&
         -5960 N^2-2112 N-288~.
\end{eqnarray}
%--------------------------------------------------------------------
Eqs.~(\ref{eq:ggq3},\ref{eq:ggg3}) confirm previous results in \cite{Vogt:2004mw}
by a first direct diagrammatic calculation, here in the massive case.

The leading singlet eigenvalue for the gluonic anomalous dimensions 
$\gamma_{gj}^{(3)},~j=q,g$ in form of
%--------------------------------------------------------------------
\begin{eqnarray}
         {\gamma}_{gg}^{(2),n_f^2}
 + \frac{{\gamma}_{gq}^{(2), n_f^2}
                \gamma_{qg}^{(0)}}
        {{\gamma}_{gg}^{(0), n_f} n_f} 
\end{eqnarray}
%--------------------------------------------------------------------
has been calculated in \cite{BennettGracey98} for the leading $n_f$ contribution,
$\propto n_f^2$. We also confirm this result by a direct massive calculation. 

Usually the calculation in $N$-space is being performed multiplying the massive OMEs
and the parton distributions analytically\footnote{For Mellin-space representations
of a wide class of parton densities see \cite{Blumlein:2011zu}.}, cf. e.g. 
\cite{Blumlein:1997em}. The corresponding analytic continuations of harmonic sums up to 
weight {\sf w=8} are given in \cite{ANCONT}.
Only a single numerical contour integral around the singularities 
has to be performed, allowing for very fast implementations.

The OMEs (\ref{eq:OME1},\ref{eq:OME2}) can also be given in $x$-space directly for
codes operating in $x$-space only. They are given by :
%--------------------------------------------------------------------
\begin{eqnarray}
A_{gq,Q}^{(3),n_fT_F^2,\MS}(x) &=&
C_F n_f T_F^2 \Biggr\{
 \left(\frac{32 x}{9}+\frac{64}{9x}-\frac{64}{9}\right)
  \ln^3\left(\frac{\bar{m}^2}{\mu^2}\right)
 +\Biggl[
     \left( -\frac{16 x}{3}-\frac{32}{3x}+\frac{32}{3} \right) \HA_1^2
\N\\ &&
    +\left(\frac{256 x}{9}+\frac{320}{9 x}-\frac{320}{9} \right) \HA_1
    +\frac{608 x}{27}+\frac{2176}{27 x}-\frac{2176}{27}
  \Biggr] \ln\left(\frac{\bar{m}^2}{\mu^2}\right)
\N\\ &&
 +\left(\frac{32 x}{27}+\frac{64}{27x}-\frac{64}{27}\right) \HA_1^3
 +\left(-\frac{256x}{27}-\frac{320}{27 x}+\frac{320}{27}\right) \HA_1^2
\N\\ &&
 +\left(\frac{256 x}{27}-\frac{128}{27x}+\frac{128}{27}\right) \HA_1
 +\left(-\frac{256x}{9}-\frac{512}{9 x}+\frac{512}{9}\right) \zeta_3
 +\frac{12608 x}{243}
\N\\ &&
+\frac{24064}{243 x}-\frac{24064}{243}
\Biggl\}
\\
%-----------------------------------
A_{gg,Q}^{(3),n_fT_F^2,\MS}(x) &=&
n_f T_F^2 \Biggl\{
 \Biggl(
   C_A \Biggl[
    -\frac{64 x^2}{27}+\frac{64
x}{27}-\frac{64}{27 (x-1)_{+}}-\frac{128}{27}+\frac{64}{27 x}
   \Biggr)
   +C_F\Biggl[
     -\frac{256 x^2}{27}
\N\\ &&
-\frac{64 x}{9}+\frac{128}{9}(1+x) \HA_0 + 
\frac{64}{9}+\frac{256}{27x}
    \Biggr]
 \Biggr] \ln^3\left(\frac{\bar{m}^2}{\mu^2}\right)
\N\\ &&
 -\frac{16}{3} C_F\delta(1-x) \ln^2\left(\frac{\bar{m}^2}{\mu^2}\right)
 +\Biggl[
   C_A \Biggl[
    -\frac{608 x^2}{27}-\frac{16}{81} (144 \zeta_2-85)
x
\N\\ &&
+\frac{32}{3}(1+x) 
\HA_0^2
-\frac{44}{3}
\delta(1-x)-\frac{16}{81} (144 \zeta_2+149)+\Biggl(-\frac{832
x^2}{27}+\frac{16 x}{27}
\N\\ &&
-\frac{800}{27}\Biggr) \HA_0 + \left(-\frac{832
x^2}{27}+\frac{208 x}{9}-\frac{176}{9}+\frac{832}{27 x}\right)
\HA_1
\N\\ &&
+\frac{256}{9}(1+x) \HA_{0,1} - \frac{2176}{81 (x-1)_{+}}+\frac{224}{27 x}
   \Biggr]
   +C_F \Biggl[
    \frac{32}{3}(1+x) \HA_0^3
\N\\
&&
+\left(-\frac{256 x^2}{9}-\frac{688
x}{9}-\frac{592}{9}\right) \HA_0^2 + \Biggl(-\frac{64
x^2}{9}-\frac{64}{9} (12 \zeta_2
+5) x-\frac{64}{9} (12 \zeta_2
\N\\ &&
-41)\Biggr) \HA_0 + \left(-\frac{512 x^2}{9}-\frac{128
x}{3}+\frac{128}{3}+\frac{512}{9 x}\right) \HA_1 \HA_0 + 
\frac{256}{3}(1+x) \HA_{0,1} \HA_0
\N\\
&&
+\left(-\frac{64 x^2}{3}-16 x+16+\frac{64}{3 
x}\right)
\HA_1^2
-\frac{20}{3} \delta(1-x) +\frac{64}{27} x^2 (18
\zeta_2-7)
+\frac{64}{9} x (3 \zeta_2
\N\\ && 
+3 \zeta_3-28)+\frac{64}{9} (6
\zeta_2+3 \zeta_3+10)+\left(-\frac{64 x^2}{9}-\frac{416x}{3}
+\frac{736}{3}-\frac{896}{9 x}\right) \HA_1
\N\\&& 
+\left(\frac{128
x^2}{9}+\frac{64 x}{3}-\frac{256}{3}-\frac{512}{9 x}\right)
\HA_{0,1}-\frac{256}{3}(1+x) \HA_{0,0,1}
+ 64 (1+x) \HA_{0,1,1}
\N\\
&& 
+\frac{3904}{27 x}
   \Biggr]
 \Biggr] \ln\left(\frac{\bar{m}^2}{\mu^2}\right)
 +C_A \zeta_3 \left(
   \frac{512 x^2}{27}-\frac{512 x}{27}+\frac{512}{27
(x-1)_{+}}+\frac{1024}{27}-\frac{512}{27 x}
  \right)
\N
\end{eqnarray}
\begin{eqnarray}
&& 
 +C_F \zeta_3
  \left(\frac{2048 x^2}{27}+\frac{512 x}{9}-\frac{1024}{9}(1+x)\HA_0
-\frac{512}{9}-\frac{2048}{27 x}
  \right)
 +C_A \Biggl[
   \frac{128}{81}(1+x)
\HA_0^3
\N\\
&& 
+\left(-\frac{208 x^2}{81}+\frac{812
x}{81}+\frac{320}{81}\right) \HA_0^2+\Biggl(-\frac{8624
x^2}{243}
-\frac{8}{81} (48 \zeta_2-199) x
-\frac{16}{27} (8
\zeta_2+19)
\N\\ && 
-\frac{64}{27 (x-1)}\Biggr) \HA_0 + \left(-\frac{416
x^2}{81}+\frac{56 x}{27}-\frac{88}{27}+\frac{416}{81 x}\right) \HA_1 \HA_0
+\left(\frac{64}{27} \zeta_2 - \frac{310}{27}\right) \delta(1-x)
\N\\ && 
+\frac{128}{27}(1+x)\HA_{0,1}
\HA_0 + \left(\frac{208 x^2}{81}-\frac{20
x}{9}+\frac{44}{27}-\frac{208}{81 x}\right) \HA_1^2
\N\\ && 
-\frac{416}{729} x^2 (9 \zeta_2+113)-\frac{8}{729}
(2088 \zeta_2-864 \zeta_3+6055)-\frac{8}{729} x (2601 \zeta_2-864
\zeta_3-4883)
\N\\ && 
+\left(-\frac{8624 x^2}{243}+\frac{2600
x}{81}-\frac{872}{81}+\frac{4592}{243 x}\right) \HA_1 + \left(\frac{832
x^2}{81}+\frac{2144 x}{81}+\frac{2120}{81}-\frac{416}{81 x}\right)
\HA_{0,1}
\N\\ && 
-\frac{128}{27}(1+x)(
\HA_{0,0,1} + \HA_{0,1,1}) - \frac{24064}{729 (x-1)_{+}}+\frac{32320}{729 x}
  \Biggr]
\N\\ && 
 +C_F \Biggl[
   \frac{32}{27}(1+x) \HA_0^4+\left(-\frac{128 x^2}{81}+\frac{256
x}{81}+\frac{64}{81}\right) \HA_0^3+\Biggl(-\frac{2176
x^2}{81}-\frac{32}{81} (18 \zeta_2+107) x
\N\\ && 
-\frac{32}{81} (18
\zeta_2-1)\Biggr) \HA_0^2+\left(-\frac{128 x^2}{27}-\frac{32
x}{9}+\frac{32}{9}+\frac{128}{27 x}\right) \HA_1
\HA_0^2+\frac{64}{9}(1+x) \HA_{0,1}
\HA_0^2
\N\\ && 
+\left(\frac{128 x^2}{27}+\frac{32
x}{9}-\frac{32}{9}-\frac{128}{27 x}\right) \HA_1^2
\HA_0+\Biggl(-\frac{128}{243} (18 \zeta_2-1) x^2
-\frac{64}{243} (333
\zeta_2-108 \zeta_3
-410) x
\N\\ && 
-\frac{64}{243} (225 \zeta_2-108
\zeta_3-1292)\Biggr) \HA_0+\left(-\frac{4352 x^2}{81}-\frac{320
x}{9}+\frac{704}{9}+\frac{896}{81 x}\right) \HA_1
\HA_0
\N\\ && 
+\left(\frac{512 x^2}{27}+\frac{2560
x}{27}+\frac{1408}{27}-\frac{256}{27 x}\right) \HA_{0,1}
\HA_0 - \frac{128}{9}(1+x)\left(\HA_{0,0,1} \HA_0
+ \HA_{0,1,1} \HA_0\right)
\N\\ && 
+\left(\frac{256 x^2}{81}+
\frac{64}{27}(x-1)-\frac{256}{81 x}\right)
\HA_1^3+\left(-\frac{1472 x^2}{81}-\frac{64}{9}(x-1)+\frac{1472}{81 x}\right) 
\HA_1^2
\N\\ && 
+\left(\frac{1024}{9}\zeta_2^2 
- \frac{1312}{81} \right) \delta(1-x)
-\frac{64}{405} x \left(63 \zeta_2^2+145 \zeta_2-120
\zeta_3+1720\right)+\frac{64}{729} x^2 (414 \zeta_2
\N\\ && 
-108
\zeta_3-1165)-\frac{64}{405} \Biggl(63 \zeta_2^2
-215 \zeta_2
-30+
\zeta_3-1675\Biggr)-\Biggl(\frac{128}{243} (18 \zeta_2-1)
x^2
\N\\
%\end{eqnarray}
%\begin{eqnarray}
&& 
+\frac{64}{27} [(3 \zeta_2+44) x-(3
\zeta_2+80)]-\frac{128 (18 \zeta_2-163)}{243 x}\Biggr)
\HA_1
\N\\
&& 
+\left(\frac{1408 x^2}{81}+\frac{128}{81} [(9 \zeta_2+37)
x+(9 \zeta_2-71)]-\frac{896}{81 x}\right)
\HA_{0,1}+\Biggl(-\frac{512 x^2}{27}-\frac{2560
x}{27}
\N\\ && 
-\frac{1408}{27}+\frac{256}{27 x}\Biggr)
\HA_{0,0,1}+\left(\frac{256 x^2}{27}+\frac{1664
x}{27}+\frac{1664}{27}+\frac{256}{27 x}\right)
\HA_{0,1,1}+\frac{128}{9} (1+x)\Biggl(\HA_{0,0,0,1}
\N\\ && 
+ \HA_{0,0,1,1} - 2 \HA_{0,1,1,1} \Biggr) +\frac{79744}{729 x}
 \Biggr]
 \Biggr\},
\end{eqnarray}
%--------------------------------------------------------------------
with the harmonic polylogarithms $\HA_{\vec{a}} \equiv \HA_{\vec{a}}(x)$ over the 
alphabet $\mathfrak{A} 
=  \{0,1,-1\}$ \cite{Remiddi:1999ew}. They can be expressed in terms of elementary 
functions and 
the 
Nielsen integrals \cite{NIELS}~: 
$\HA_0(x) = \ln(x),
 \HA_1(x) = -\ln(1-x),
 \HA_{0,1}(x) = \Li_2(x), 
 \HA_{0,0,1}(x) = \Li_3(x), 
 \HA_{0,1,1}(x) = {\rm S}_{1,2}(x), 
 \HA_{0,0,0,1}(x) = \Li_4(x), 
 \HA_{0,0,1,1}(x) = {\rm S}_{2,2}(x)$ and 
$\HA_{0,1,1,1}(x) = {\rm S}_{1,3}(x)$, with 
%--------------------------------------------------------------------
\begin{eqnarray}
S_{n,p}(x) &=& \frac{(-1)^{(n+p-1)}}{(n-1)!p!} \int_0^1 \frac{dy}{y} \ln^{(n-1)}(y)
\ln^p(1-x y)~,\\
\Li_n(x) &=& S_{n-1,n}(x)~.
\end{eqnarray}
%--------------------------------------------------------------------
Here $\Li_n(x)$ denotes the polylogarithm. All higher functions but ${\rm S}_{2,2}(x)$
can be reduced to polylogarithms by the argument relation $x \rightarrow (1-x)$.
Numerical implementations of the functions $S_{n,p}(x)$ were given in 
\cite{Kolbig:1969zza}.

At small values of $x$ the functions
$A_{gq(g),Q}^{(3),n_fT_F^2,\MS}(x)$ are singular as $\propto 1/x$, or in 
$N$-space like $\propto 1/(N-1)$, unlike the quarkonic contributions given in 
\cite{ABKSW11NPB} with a leading pole $\propto 1/N$. One notices that the number of
functions needed in $x$-space to express $A_{gq(g),Q}^{(3),n_fT_F^2,\MS}$ 
is larger than in $N$-space, as has been found also in other analyses, cf. 
\cite{BBK07NPB,Bierenbaum:2007pn,Blumlein:2009tj}, requesting very careful 
numeric implementations.
%%%%%%%%%%%%%%%%%%%%%%%%%%%%%%%%%%%%%%%%%%%%%%%%%%%%%%%%%%%%%%%%%%%%%%%
\section{Conclusions}
\label{sec:4}
%%%%%%%%%%%%%%%%%%%%%%%%%%%%%%%%%%%%%%%%%%%%%%%%%%%%%%%%%%%%%%%%%%%%%%%

\vspace{1mm}
\noindent
We have calculated the contributions $O(\alpha_s^3 n_f T_F^2 C_{A,F})$ to the massive
OMEs with local operator insertions on gluonic lines and veritces at general values 
of the Mellin variable $N$. These matrix elements are needed to describe the transition 
functions in the VFNS. In the calculation representations of the Feynman diagrams by
generalized hypergeometric functions play an essential role. They allow the $\ep$-expansion 
into nested sums, which can be solved using modern summation technologies. The number of 
these sums is very large, although their structures exhibit similarities. One may synchronize
these sums, leading to a low number, however, with voluminous intermediate terms. The solution 
of the latter sums turns out to be more economic. The final results in $N$ space can be expressed
by rational functions in $N$ and single harmonic sums up to $S_3(N)$. We also derived the 
corresponding $x$-space results, which have a more involved structure and depend on  
six Nielsen integrals.

%---------------------------------------------------------------------------------------------------
\vspace*{2mm}
\noindent
{\bf Acknowledgment.}
For discussions we would like to thank J. Ablinger, A. De Freitas, and F.~Wi\ss{}brock.
This work has been supported in part by DFG Sonderforschungsbereich Transregio 9, 
Computergest\"utzte Theoretische Teilchenphysik, Austrian Science  Fund (FWF) grant 
P203477-N18, and EU Network {\sf LHCPHENOnet} PITN-GA-2010-264564.
%---------------------------------------------------------------------------------------------------

\end{document}